\documentclass{WileyMSP-template}

\begin{document}

\title{\textbf{Coupled resonator acoustic waveguides-based acoustic interferometers designed within two-di\-men\-si\-o\-nal phononic crystals: experiment and theory}}

\maketitle

\author{David Mart\'inez-Esquivel}
\author{Rafael Alberto M\'endez-S\'anchez}
\author{Hyeonu Heo}
\author{Angel Marbel Mart\'inez-Arg\"uello*}
\author{Miguel Mayorga-Rojas}
\author{Arup Neogi*}
\author{Delfino Reyes-Contreras*}

\begin{affiliations}
D. Mart\'inez.\\
Programa de Doctorado en Ciencias, Facultad de Ciencias, Universidad Aut\'onoma del Estado de M\'e\-xi\-co, Campus ``El Cerrillo", Toluca, Estado de M\'exico, M\'exico\\

R. A. M\'endez-S\'anchez\\
Instituto de Ciencias F\'isicas, Universidad Nacional Aut\'onoma de M\'exico, Cuernavaca, Morelos,\\
M\'exico\\

H. Heo\\
Graduate Program in Acoustics, The Pennsylvania State University, University Park, Pennsylvania 16802, USA\\

A. M. Mart\'inez-Arg\"uello\\
Instituto de F\'isica, Benem\'erita Universidad Aut\'onoma de Puebla, C. P. 72570 Puebla, Pue., M\'exico\\
*E-mail: angelm@ifuap.buap.mx

M. Mayorga\\
Facultad de Ciencias, Universidad Aut\'onoma del Estado de M\'exico, Campus ``El Cerrillo", Toluca, Estado de M\'exico, M\'exico\\

A. Neogi\\
Institute of Fundamental and Frontier Sciences, University of Electronic Science and Technology of Chi\-na, Chengdu, 610054, China\\
*E-mail: arup@uestc.edu.cn

D. Reyes\\
Laboratorio de Ac\'ustica y Nanomateriales (LAN), Facultad de Ciencias, Universidad Aut\'onoma del Estado de M\'exico, Campus "El Cerrillo", Toluca, Estado de M\'exico, M\'exico\\
*E-mail: dreyesc@uaemex.mx

\end{affiliations}


\begin{abstract}

The acoustic response of defect-based acoustic interferometer-like designs, known as Coupled Resonator Acoustic Waveguides (CRAWs), in two-dimensional phononic crystals (PnCs) is reported. The PnC is composed of steel cylinders arranged in a square lattice within a water matrix with defects induced by selectively removing cylinders to create Mach-Zehnder-like (MZ) defect-based interferometers. Two defect-based acoustic interferometers of MZ-type are fabricated, one with arms oriented horizontally and another one with arms oriented diagonally, and their transmission features are experimentally characterized using ultrasonic spectroscopy. The experimental data are compared with finite element method (FEM) simulations and with tight-binding (TB) calculations in which each defect is treated as a resonator coupled to its neighboring ones. Significantly, the results exhibit excellent agreement indicating the reliability of the proposed approach. This comprehensive match is of paramount importance for accurately predicting and optimizing resonant modes supported by defect arrays, thus enabling the tailoring of phononic structures and defect-based waveguides to meet specific requirements. This successful implementation of FEM and TB calculations in investigating CRAWs systems within phononic crystals paves the way for designing advanced acoustic devices with desired functionalities for various practical applications, demonstrating the application of solid-state electronics principles to underwater acoustic devices description.

\end{abstract}

\section{Introduction}
Defects engineering in phononic crystals (PnCs) has convincingly demonstrated the feasibility of guiding waves within their acoustic bandgap when defects are optimally induced~\cite{Pennec2010,Shelke2014,Reyes2019}. PnCs, also known as acoustic bandgap materials, consist of sound scatterers periodically arranged in a matrix in which both components exhibit different physical properties such as mass density, speed of sound, and Young's modulus. This spatially periodic mass distribution enables the observation of acoustic or elastic bandgaps. Similar to stopbands in electronic or photonic crystals, phononic bandgaps are frequency intervals where wave propagation is forbidden across the crystal. By introducing defects that break the crystal symmetry, typically induced by removing scatterers, defect modes become allowed within the bandgap~\cite{Reyes2019,Laude2021,Reyes2020,Ghasemi2016}. While a single defect acts as a resonator, promoting a unique in-gap eigenmode, a group of defects gives rise to multiple coupled eigenmodes leading to miniband transmission within the stopband interval~\cite{Escalante2013}.

The coupling of resonant modes in defect-based PnCs has been successfully described approximating these as Coupled Resonator Acoustic Waveguides (CRAWs) within the framework of tight-binding models (TBM)~\cite{Wang2018a,Wang2023,Ji2022,Sainidou2006,Cicek2016,Ramirez2020}. TBM, widely employed in solid-state physics to study the electronic properties of materials, can be extended to investigate other excitations, such as phonons which represent the quantized vibrations of crystal lattices~\cite{Escalante2013,Ramirez2020}. In this approach, a PnC-based CRAW is discretized into separate sites and the dynamics of phonons are described by considering the interactions between neighboring sites. The propagation of waves in CRAWs through the TBM approximation can be analyzed by solving the equations of motion for phonons in the crystal lattice and considering the interaction strengths between adjacent lattice sites~\cite{Ramirez2020,Kaina2017,AMMA2022}. The dynamical matrix of the phononic crystal, which incorporates the interactions between neighboring sites, provides information about the allowed phononic modes~\cite{AMMA2022,Mendez2021}.

CRAWs engineered within phononic crystals consist of multiple waveguide segments coupled through resonators or defects forming a structure of interconnected unit cells. Each unit cell contains the waveguide segment and resonator or defect composing the CRAW~\cite{Laude2021,kadmiri2020}. These structures enable precise manipulation and control of acoustic waves, facilitating the creation of waveguides with desirable transmission properties and the formation of localized modes within band gaps. The theoretical framework based on TB theory allows for constructing the Hamiltonian that describes the dynamics of acoustic waves propagating along these coupled resonant waveguides. The Hamiltonian incorporates kinetic terms representing the wave´s energy in the waveguide segment and potential terms accounting for the coupling between adjacent segments through the resonators or defects~\cite{Ramirez2020,AMMA2022}. This approach facilitates determining the coupling strength between the waveguide segments which directly impacts the transmission properties and formation of localized modes~\cite{Escalante2013}. As a result, it becomes feasible to predict the resonant frequencies of the coupled waveguide structure. These resonant frequencies are associated with the formation of localized modes representing waveguide modes confined within the resonator or defect regions~\cite{Mendez2021,Ma2022}.

The application of the TB model in the design and optimization of CRAWs systems in PnCs with specific functionalities such as enhanced transmission within certain frequency ranges, strong localization of waveguide modes, and efficient filtering of acoustic waves, is experiencing significant growth in the field of acoustic and mechanical phononic crystal structures~\cite{Ma2022,Betancur2023,Sara2023,Chen2020,Ma2021,Xue2021,Choi2021}. Besides that, 1D and 2D Phononic meta-materials can be also used to manipulate acoustic wave propagation, neither modeling 1D phononic structures with hyperbolic dispersion ~\cite{Zubov2020,Jin2020}, nor 2D phononic structures with tailorable equifrequency contour for focusing [c], or beam steering ~\cite{Walker2020,Heo2023}.

This study presents the adaptation of the TBM to describe the acoustic response of defect-based acoustic interferometer-like designs, referred to as CRAWs systems in two-dimensional PnCs. The phononic crystal is composed of steel cylinders, arranged squarely in a water matrix, with defects induced by removing specific cylinders to create Mach-Zehnder-like defects-based interferometers. Two defect-based acoustic interferometers of MZ-type are constructed, one with arms oriented horizontally and another one with arms oriented diagonally, and their transmission features are experimentally measured using ultrasonic spectroscopy. The tight-binding approximation is applied to both models considering each defect as a resonator coupled to its neighboring ones. Finite element method (FEM) simulations and bandstructure calculations based on the linearized Navi\`er-Stokes equation using the COMSOL platform are also considered. The obtained results are compared with the experimental data, demonstrating the application of solid-state electronics principles to underwater acoustic-devices analysis. Remarkably, our results show excellent agreement with the experimental and simulation data. This comprehensive match is crucial for predicting and optimizing resonant modes supported by defect arrays, enabling the customization of phononic structures and defect-based waveguides to specific requirements. 

\section{Results}

\subsection{Ultrasonic spectroscopy results}
\begin{figure}
\centering
\includegraphics[width=\linewidth]{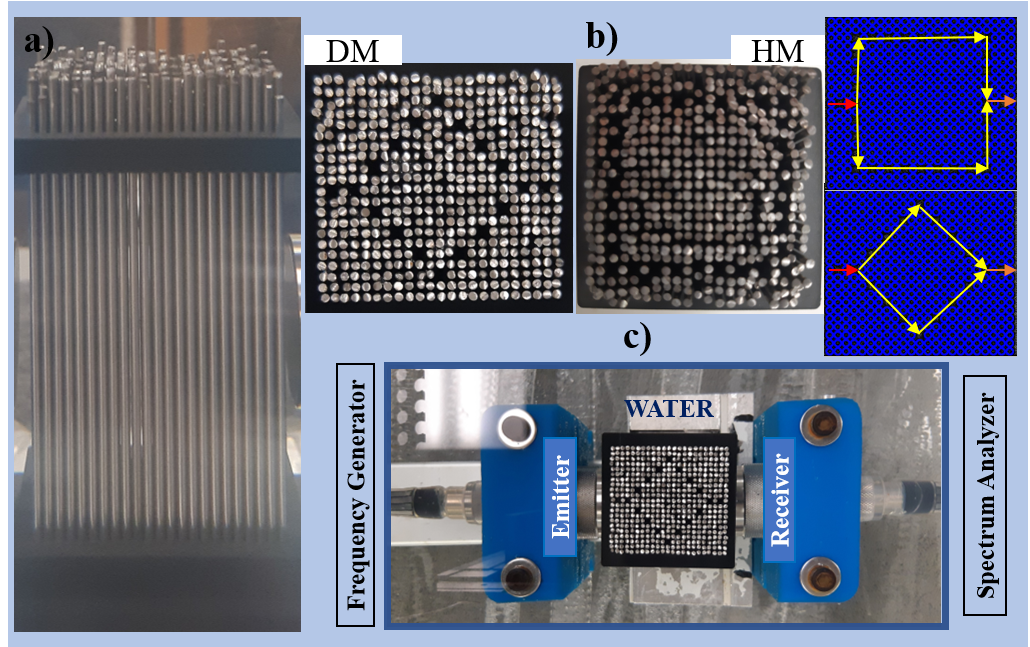}
\caption{(Color online) Actual interferometers and experimental setup for ultrasonic spectroscopy analysis. a) Lateral-actual view of the perfect PnC. b) Top-actual view of the DM (left) and HM (center) interferometers, (right)-schematic of both designs showing the defect paths. c) Ultrasonic spectroscopy setup.}
\label{fig:ExpModels}
\end{figure}

A phononic crystal (PnC) composed of a $23\times 21$ arrangement of stainless-steel cylinders of radius 0.8 mm is immersed in a water matrix with a lattice parameter of 2 mm, as depicted in Figure~\ref{fig:ExpModels} a). Two defect-based interferometers, inspired by Mach-Zehnder-type designs, as shown in Figure~\ref{fig:ExpModels} b) are constructed. Those are labeled as the diagonal model (DM) and the horizontal model (HM), placed on the left and center side of Figure~\ref{fig:ExpModels} b), respectively. The DM-interferometer consists of a defect input and two arms built with defects diagonally arranged, which coincide with another defect, as illustrated on the left side of Figure~\ref{fig:ExpModels} b). The HM-interferometer has one input defect and two linear arms, as observed on the center side of Figure~\ref{fig:ExpModels} b). The right side of Figure~\ref{fig:ExpModels} b) are schematic representations of the DM and HM interferometers; red and orange arrows describe the input and output points, respectively, and the yellow ones are the defect-paths within the phononic structure. To maintain their positions without perturbation, all cylinders are stacked on perforated ABS plastic bases. Experimental results have demonstrated that groups of defects continuously induced in a phononic crystal can function as waveguides for waves whose frequencies fall within their phononic bandgap~\cite{Reyes2019,Reyes2020}.

\begin{figure}
\centering
\includegraphics[width=\linewidth]{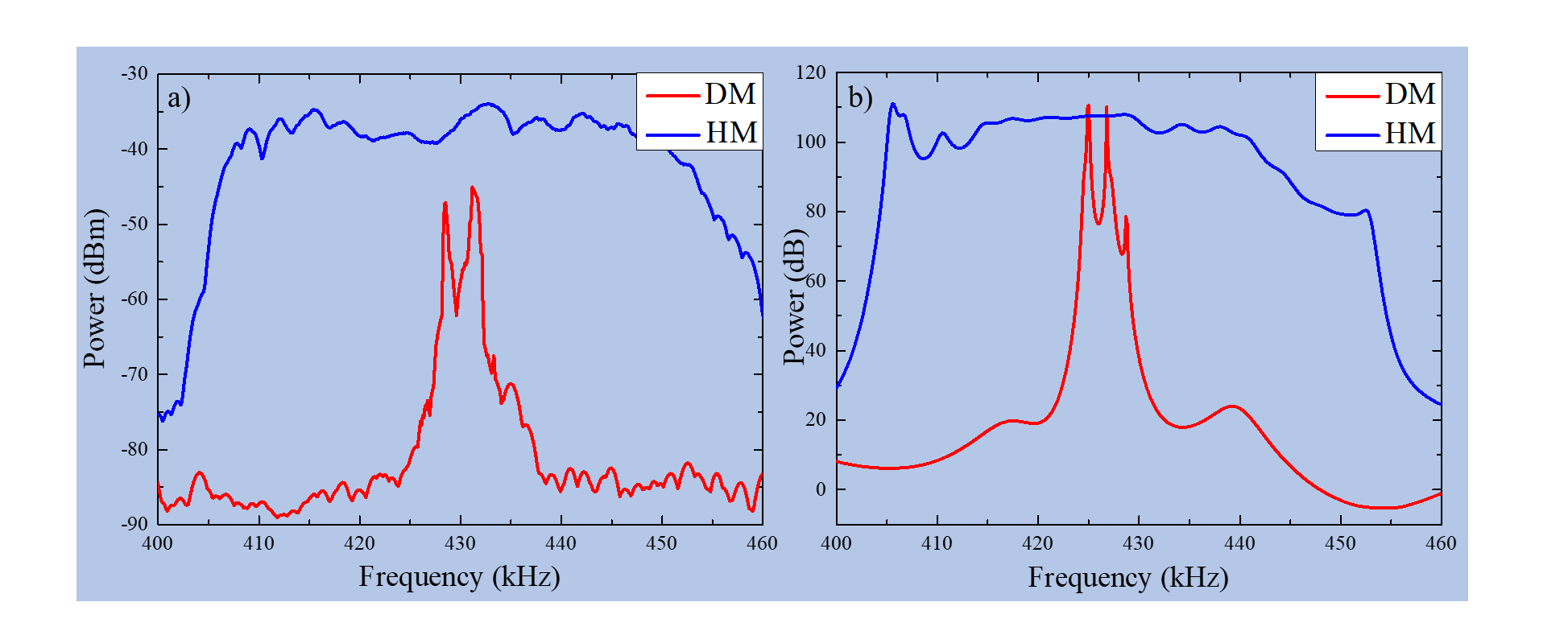}
\caption{(Color online) Experimental and theoretical (FEM) transmission characterization for the designed interferometers. a) Experimental transmission spectra for both DM (blue line) and HM (red line) interferometers: inset corresponds to the FEM simulated results. b) FEM spectra from the COMSOL platform for the two models.}
\label{fig:CRAWsExperiment}
\end{figure}

The experimental transmission spectra for both designed defect-based acoustic interferometers, DM and HM, are depicted in Figure~\ref{fig:CRAWsExperiment} a) in red and blue lines, respectively, covering the frequency range from 400 to 460 kHz. \textbf{Acoustic response was experimentally recorded in dBm units, which have its logarithmic equivalence in watt (+20 dBm (0.1 W)), and represent power quantity. It was selected to use this unit, which is related to the wave intensity when considering the transducer area, by simplicity as the raw information from the spectrum analyzer. It is evident that the transmitted acoustic power of the models differs significantly depending on the spatial disposition of the defects forming each interferometer, which yields different transmitted power at the same frequencies.} It has been reported the perfect crystal exhibits a full bandgap from approximately 380 to 480 kHz, as will be shown below, and previously reported by our group~\cite{Reyes2019,Reyes2020,Walker2020}.

%

For the DM array, a narrower transmission band is induced within the acoustic gap compared to the response recorded for the HM. This behavior indicates the selective transmission of waves through the first model. The observed differences in the acoustic response between both models are due to the spatial configuration of the defects, as it has been previously demonstrated that nine defects configured diagonally allow for a single peak approximately 16 times narrower ($\sim 426$ - 429 kHz) than when defects are arranged horizontally (relative to the propagation axis), which results in multiple peaks ($\sim 405$ - 460 kHz) ~\cite{Reyes2020}. The DM interferometer exhibits a miniband with at least two well-defined in-gap resonances at 427.4 and 431.2 kHz along with a smaller one at 434.8 kHz, see the red line in Figure~\ref{fig:CRAWsExperiment} a). In contrast, multiple resonant peaks were recorded for the HM design, resulting in a new well-defined, and broader in-gap transmission band. These multiple resonant peaks arise from the collectively coupled resonant modes supported by each individual defect~\cite{Reyes2019,Reyes2020}.  

%
\begin{figure}
\centering
\includegraphics[width=0.8\linewidth]{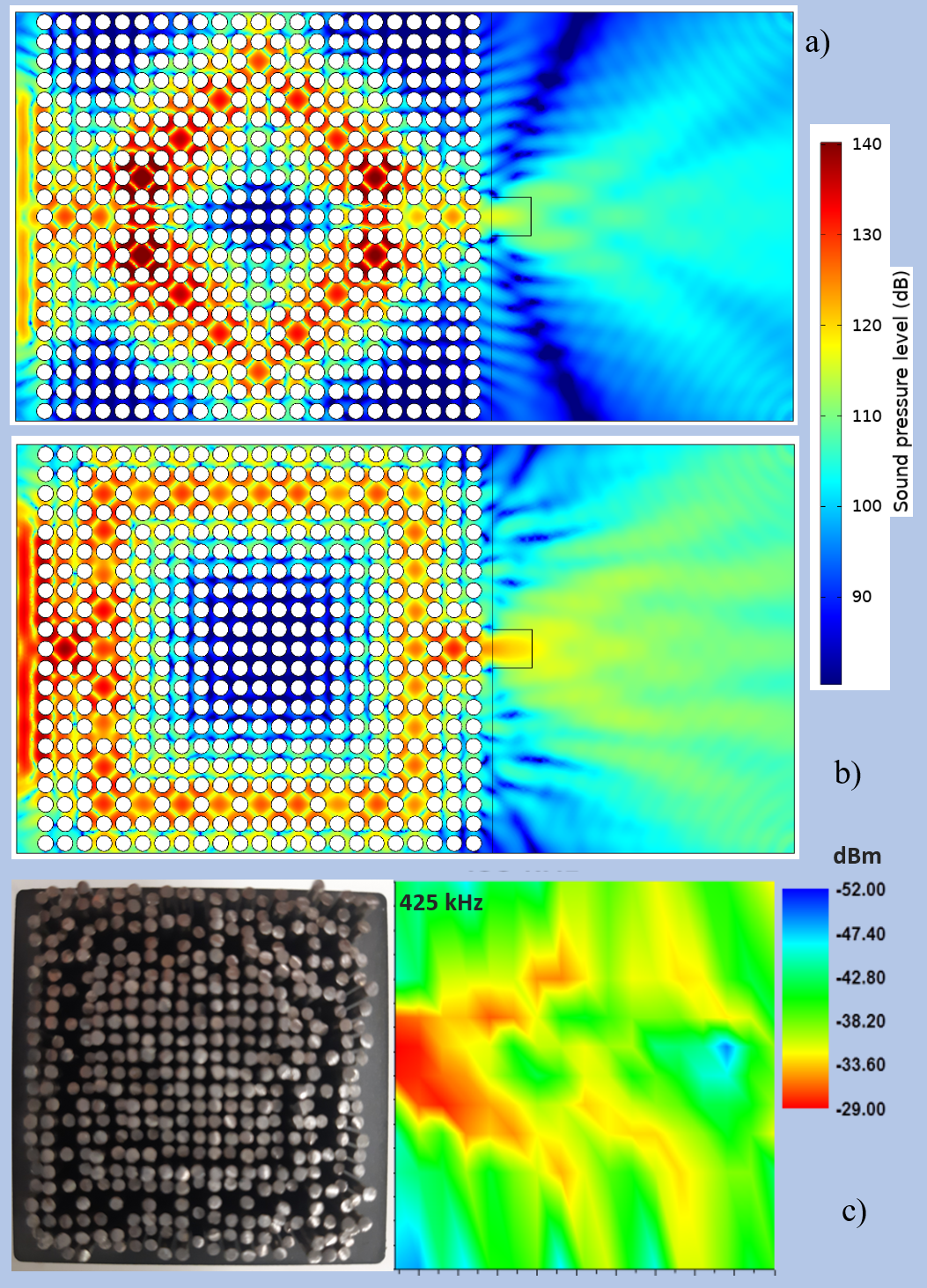}
\caption{(Color online) Pressure maps for the DM and HM interferometers. Obtained from COMSOL for the a) DM and b) HM interferometers for the frequency of 425.0 kHz. As observed, the acoustic energy is being transported across the defect paths. c) Experimental pressure map for the HM model, where the acoustic energy distribution has a similar distribution as for the FEM results.}
\label{fig:CRAWsPressure}
\end{figure}

\subsection{FEM Calculations}
The experimentally analyzed PnCs structures are designed using the COMSOL Multiphysics platform (version 5.3), and their acoustic properties are calculated employing the Finite Element Method (FEM) which serves as the foundation of the cited platform. The 2D phononic structures consist of stainless-steel scatterers of radius 0.8 mm arranged in a $23\times 21$ square lattice within a water matrix with a lattice parameter of 2 mm. The density and the longitudinal sound velocity values for stainless steel and water are obtained from the library included in the COMSOL platform: $\rho_{_\mathrm{S}} = 7800$ kg/$\mathrm{m}^{3}$, $\rho_{_\mathrm{W}} = 1000$ kg/$\mathrm{m}^{3}$, \textbf{$c_{_\mathrm{S}} = 5790$ m/$\mathrm{s}$, and $c_{_\mathrm{W}} = 1480$ m/$\mathrm{s}$}, all of them at 20${}^{\circ}\mathrm{C}$. For water and steel, we used the density and longitudinal speed of sound. Additionally, Young's modulus and Poisson ratio were incorporated for the solid steel, 205 GPa and 0.28, respectively, which are demandant for the simulations and were taken from the library. The simulations are obtained for the same frequency range as that of the experimental ones. The DM and HM interferometers are depicted in Figure~\ref{fig:TBmodels}, panels a) and b) respectively. The acoustic-solid interaction model is utilized to simulate the acoustic response of each model with a frequency sweep using a plane wave radiation source. Figure ~\ref{fig:CRAWsExperiment} b) corresponds to the FEM spectra, which nearly agree with the experimental results recorded through ultrasonic spectroscopy (Figure~\ref{fig:CRAWsExperiment} a)). For the DM interferometer three numerical peaks were found at 425, 426.8, and 429 kHz, and are associated with the experimentally recorded peaks. \textbf{It could be convenient to clarify that in COMSOL, the input signal is given in Pa, which is a unit of pressure that is equivalent to 93.9794 dB Sound Pressure Level (SPL), and that is the reason because the transmitted signal is in dB, representing the SPL.} 

Additionally, pressure maps were generated to gain insight into the acoustic energy distribution inside and outside the phononic structures. Figure~\ref{fig:CRAWsPressure} a) shows the pressure map for the DM interferometer at a frequency of 425.0 kHz which can be associated with the experimental in-gap resonant peak observed at 428.4 kHz. In the simulations, the source is placed at a distance of two millimeters from the crystal which is consistent with the experimental setup and has a size of 25 mm, corresponding to the actual transducer size. Figure~\ref{fig:CRAWsPressure} b) contains the pressure map for the HM model at the same frequency, 425.0 kHz. \textbf{These pressure maps are represented as SPL in dB, which has been clarified above.} Pressure naps provide valuable insights into the behavior of the acoustic energy for this frequency in each model, revealing how the waves follow the designed defect paths to guide them. This can be visually observed as the sound pressure at the center of both models has the lowest levels (blue color), which implies that no wave propagation at these frequencies occurs due to the bandgap. The largest sound pressure levels (red color) are observed at the input and output points and along the defect paths due to the defect array being a CRAW by itself working as a waveguide to allow transmission at the in-gap frequencies. 

To obtain an insight into the acoustic pressure distribution in the actual models, an experimental pressure map was obtained for the HM model based on the temporal evolution and equipment requirements (see experimental section). The experimental pressure map for this model is displayed in Figure~\ref{fig:CRAWsPressure} c), where the acoustic pressure in dBm units has similar characteristics as that observed in Figure~\ref{fig:CRAWsPressure} b). The sound pressure was measured 2mm away from the transducer facet. It was -17.86 dBm for a 20 Vpp from the frequency generator, implying a power loss of over ten dBm. \textbf{Here, 20 Vpp means the sinusoidal wave amplitude, which induces an input power (2mm in front of the transducer used as emitter) measured in the Spectrum analyzer of -17.86 dBm, which reduces to -29 dBm at the output point of the HM model, indicating a reduction in the experimentally measured power over 10 dB. This reduction in power is higher in the rest of the selected area, due to the green color meaning -40 dBm.} The local acoustic energy measured using a needle hydrophone as a receiver drops in intensity compared to the global pressure recorded using a transducer (Figure~\ref{fig:CRAWsExperiment} a)). However, it is still detectable by the used equipment, which means it is viable for practical measurements in actual devices or possible applications. 

%
\begin{figure}
\centering
\includegraphics[width=\linewidth]{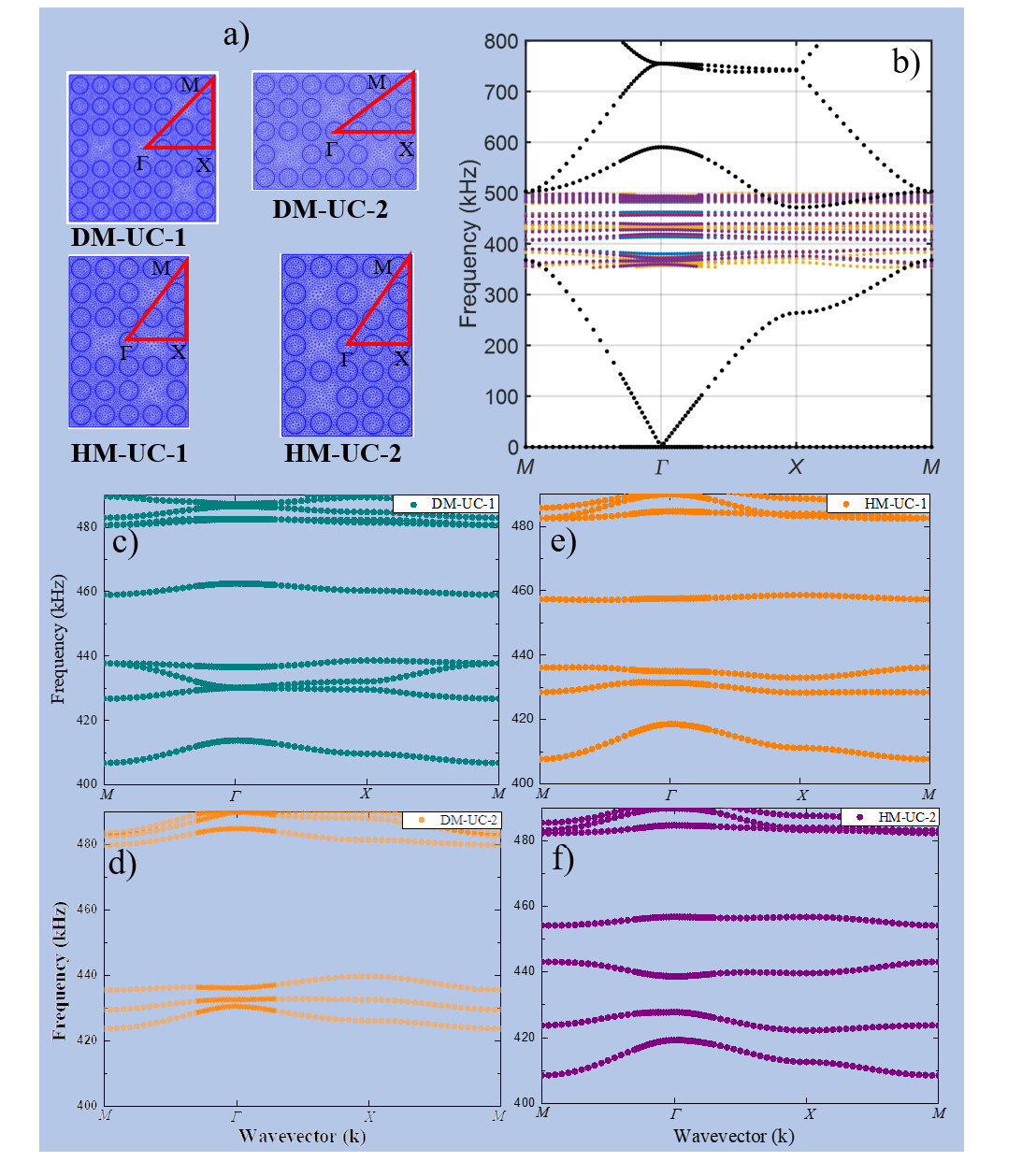}
\caption{(Color online) Band structure calculations for the selected unit cells. a) Unit cells where the k-vectors are sketched, b) Band structure for the perfect phononic crystal, where the corresponding unit cells are included. Zoomed bandstructure diagrams for the c) DM-UC-1, d) DM-UC-2, e) HM-UC-1, and f) HM-UC-2 unit cells for better visualization.}
\label{fig:BSCalculations}
\end{figure}

\subsection{Band structure calculations}
Figure~\ref{fig:BSCalculations} contains the bandstructure calculations performed to understand and correlate the observed results. For both DM and HM interferometers, two separate unit cells (UC) containing a group of defects were calculated and assigned as UC-1 and UC-2, as shown in Figure~\ref{fig:BSCalculations} a). The band structure of each interferometer was calculated utilizing Pressure Acoustics in conjunction with Solid Mechanics and applying the Floquet periodic boundary condition. The band structure was calculated employing a wavevector normalized by the period of the crystal. Specifically, we defined the X and M points within the Brillouin Zone (BZ) as illustrated in ~\ref{fig:BSCalculations} a). Notably, we established the center of the superlattice as $\Gamma$, and the definitions of the X and M points are in accordance with the C4 symmetry axis. In Figure~\ref{fig:BSCalculations} b), the black squares represent the calculated band structure for the perfect phononic crystal between 1 - 800 kHz. The other points in Figure~\ref{fig:BSCalculations} b) correspond to each model's unit cells, which are zoomed in Figures~\ref{fig:BSCalculations} c)-f). 

For the two-unit cells of each interferometer, DM, and HM array, in-gap solutions were obtained in the 400-460 kHz frequency interval. For the PnC, it has been reported that one defect induces an alone resonant defect mode at 427 kHz ~\cite{Reyes2020}, while several aligned defects induce defect modes whose number is exactly the number of defects; the diagonal disposition of 5 defects induced a narrow band 4 kHz-width centered at around 427 kHz. It is hard to capture the entire coupling of all the induced resonant modes due to the presence of the 18 or 34 defects composing both, DM and HM models, respectively. In the band structure diagram of the four UC, an in-gap solution in the frequency range of 424-429 kHz was recorded, which can be related to the fact that at each of these, there is diagonal coupling between at least two defects. As observed in Figures~\ref{fig:BSCalculations} c), d), and e), two or three allowed modes appear between 424-238 kHz, indicating the diagonal coupling is dominating. This is more visible in Figures~\ref{fig:BSCalculations} d), where only modes in this frequency interval were obtained, which agrees with the unit cell DM-UC-2 (Figure~\ref{fig:BSCalculations} a)), where only diagonal coupling can be assigned; the inter-period separation between the first and last defect is three, which reduce the coupling strength.

However, for the HM, in both unit cells the allowed modes are separated, covering the entire range of 400-460 kHz, as seen in Figure~\ref{fig:BSCalculations} f). The modes near 410 and 460 kHz in Figures~\ref{fig:BSCalculations} c), and e) and f) could be the result of the coupling between the two horizontal continuous defects, which are present in these three unit cells. Even when the band structure diagrams do not faithfully reproduce results in Figure~\ref{fig:CRAWsExperiment}, the in-gap induced modes for the selected unit cells agree with the fact that the induced transmission band due to the HM array extends along a larger frequency interval compared with the DM model. Larger unit cells must be necessary to fully reproduce the entire behavior.

%
\begin{figure}
\centering
\includegraphics[width=0.7\linewidth]{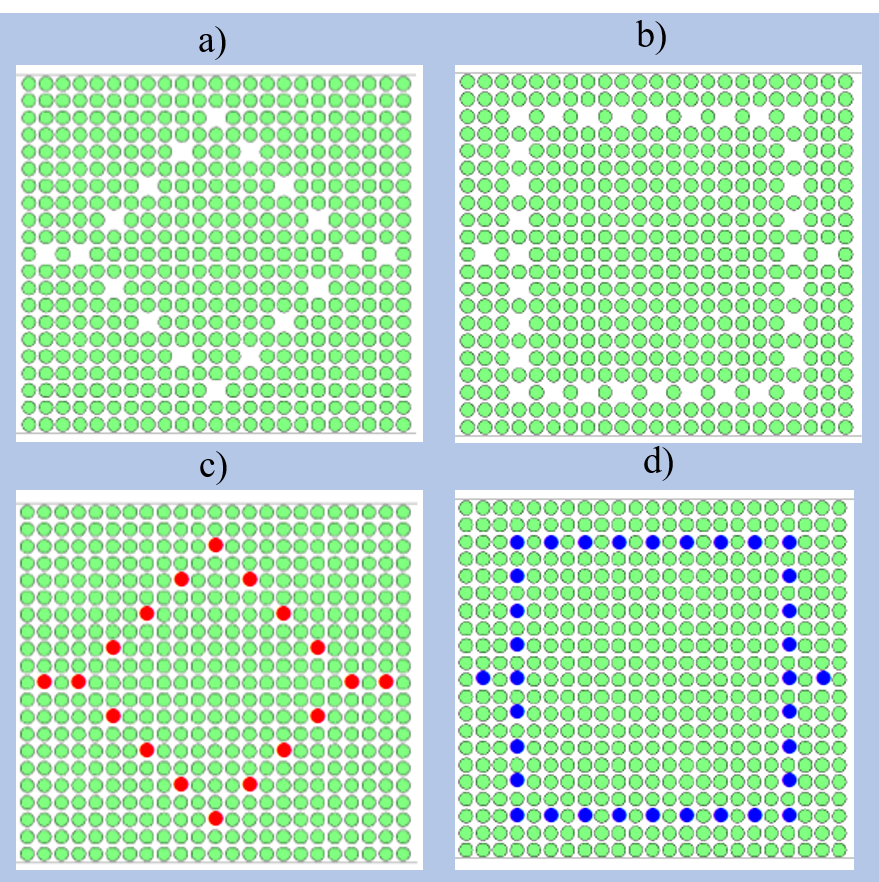}
\caption{(Color online) Schematic top-view of the models under study. The unoccupied sites represent defect cavities in the array and define a) the DM and b) the HM interferometer. These models can effectively be described by a tight-binding model defined by the red and blue dots for c) the DM and d) the HM models.}
\label{fig:TBmodels}
\end{figure}

\subsection{Tight Binding Model}
As previously mentioned, the tight-binding approximation is employed to theoretically describe the experimental and FEM simulation results for the designed CRAWs, which constitute the defects-based Mach-Zehnder-type acoustic spectrometer. The schematic representation of the models, consisting of steel cylinder arrays (top view) is illustrated in Figure~\ref{fig:TBmodels}. The unoccupied sites within that two-di\-men\-si\-o\-nal locally periodic structure or finite PnC correspond to defect cavities or impurities that define the diagonal and horizontal models shown in panels a) and b), respectively. These defect cavities are intentionally designed such that their corresponding resonant frequencies lie within the bandgap of the finite PnC, thereby allowing the localization of the resonant modes associated with the respective impurities.

The weak coupling between neighboring resonators, via evanescent Bloch waves, yields the nearest-neighbor tight-binding approximation usually found in solid state physics~\cite{AshcroftBook}. Note that this description is the same found in the coupled-resonator optical~\cite{Yariv1999,Ozbay2002}, acoustical~\cite{Reyes2019,Reyes2020,Sainidou2006,Cicek2016,Lemoult2016,Taleb2021}, and elastic~\cite{Escalante2013,Ramirez2020,AMMA2022,Mendez2021,Wang2020,Wang2018b} waveguides. Panels c) and d) in Figure~\ref{fig:TBmodels}, show a schematic of the tight-binding model of the corresponding system defined by the red and blue dots for the diagonal and horizontal model respectively. These models can be described by the following nearest-neighbor tight-binding Hamiltonian
\begin{eqnarray}
\label{eq:Hamiltonian}
H & = & \sum_{n,m} f_{n,m} | n, m \rangle \langle n, m | \nonumber \\ 
& + & \sum_{n,m} \big[ \nu_{(n,m),(n+1,m)} | n, m \rangle \langle n+1, m |  \big. \nonumber \\
& & + \, \nu_{(n,m),(n-1,m)} | n, m \rangle \langle n-1, m | \nonumber \\
& & + \, \nu_{(n,m),(n,m+1)} | n, m \rangle \langle n, m+1 | \nonumber \\
& & \big. + \nu_{(n,m),(n,m-1)} | n, m \rangle \langle n, m-1 | \big] , \nonumber \\
\end{eqnarray}
where $f_{n,m}$ is the resonance frequency of the defect cavitie at positions $| n, m \rangle$ and $\nu_{i,j}$ is the hopping frequency to nearest neighbours.


\subsubsection{Scattering matrix approach}
\label{subsec:Smatrix}

The isolated diagonal and horizontal samples of Figure~\ref{fig:TBmodels} c) and d) can be opened by attaching to them two semi-infinite single-mode perfect leads with coupling strength $\gamma^{\mathrm{L,R}}$ to the left (L) and right (R) sites, respectively. The $2\times 2$ scattering matrix, $S$-matrix, can be written as~\cite{Mahaux1969,DattaBook}
\begin{equation}
S(f) =\left(
\begin{array}{ccc}
r & t' \\
t & r'
\end{array} 
\right)
= \textbf{1}_{2} - 2\mathrm{i}\, \sin(k) W^{T}(f - H_\mathrm{eff})^{-1}W,
\label{eq:Smatrix}
\end{equation}
where $r$($r'$) and $t$($t'$) are the reflection and transmission amplitudes when the incidence is from the left (right), $\textbf{1}_{2}$ is the unit matrix of dimension $2$, $k = \arccos(f/2)$ is the wave vector at frequency $f$ supported in the leads, and $H_\mathrm{eff}$ is the effective non-Hermitian Hamiltonian given by
\begin{equation}
\label{eq:Heff}
H_\mathrm{eff} = H - \frac{\mathrm{e}^{\mathrm{i}k}}{2} WW^{T},
\end{equation}
where $H$ is the $N\times N$ Hamiltonian matrix that describes the sample with $N$ resonant states, see Eq.~(\ref{eq:Hamiltonian}). In equations~(\ref{eq:Smatrix}) and (\ref{eq:Heff}), the matrix $W$, where the superscript $T$ indicates the matrix transposition operation, is an $N\times 2$ matrix that couples the $N$ resonant states of the closed sample to the $2$ propagating modes in the leads. Its elements are defined by
\begin{equation}
W_{ij} = 2\pi \sum_{c=\mathrm{L,R}} A_{i}^{c}(f) A_{j}^{c}(f)
\end{equation}
with the coupling amplitudes given by
\begin{equation}
A_{i}^{\mathrm{L,R}}(f) = \sqrt{\frac{\gamma^{\mathrm{L,R}}}{\pi}} \left( 1 - \frac{f^2}{4} \right)^{1/4} \big( \delta_{i,\mathrm{L}} + \delta_{i, \mathrm{R}} \big) 
\label{eq:amplitudes}
\end{equation}
with $\delta_{i,j}$ the usual Kronecker delta. Furthermore, the frequency dependence in $H_\mathrm{eff}$ can be neglected since $\arccos(f/2)$ changes slightly at the center of the band. Then, from the two-channel $S(f)$-matrix of equation~(\ref{eq:Smatrix}), the transmission is obtained from the transmission amplitude as $|t|^{2}$.

%
\begin{figure}
\centering
\includegraphics[width=0.8\linewidth]{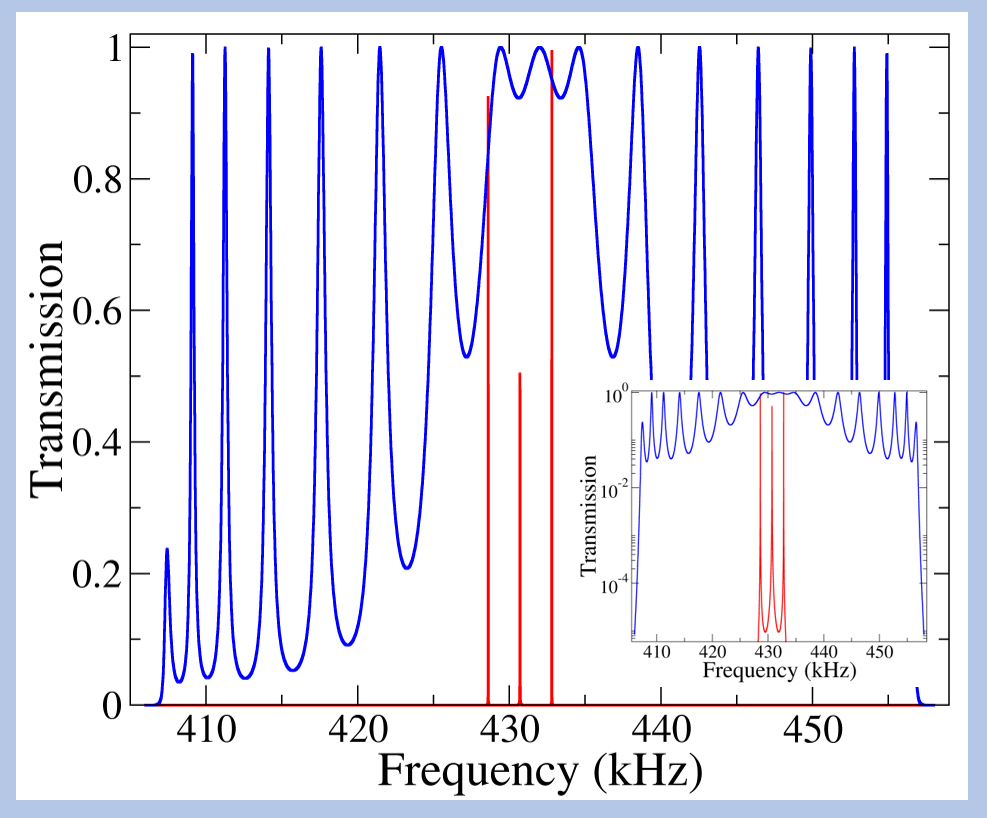}
\caption{(Color online) Transmission as a function of frequency obtained from the tight-binding model for both the diagonal and horizontal model in red and blue solid lines, respectively. The inset shows the transmission in the frequency range of interest.}
\label{fig:CRAWsTB}
\end{figure}
%


%
\textbf{
\begin{table*}
\begin{center}
\begin{tabular}{clllll}
\hline
Interferometer & $f_{0}$ (kHz) & $\nu$ (kHz) & $\nu'$ (kHz) & $\gamma$ (kHz)  \\[0.5ex]
\hline
Horizontal & 432.0 & 12.0 & 12.0 & 10.0 \\
\hline
Diagonal & 430.7 & 57.0 & 4.2 & 0.005  \\
\hline
\end{tabular}
\end{center}
\caption{\label{tab:parameters}
Parameters for constructing Hamiltonian~(\ref{eq:Hamiltonian}) and $S$-matrix~(\ref{eq:Smatrix}) along with equations~(\ref{eq:Heff})-(\ref{eq:amplitudes}) for both the horizontal and diagonal interferometer.}
\end{table*}
}

The tight-binding model and the scattering approach, as described above, are applied to elucidate the behavior of both interferometers. For this purpose, further simplifications of Hamiltonian of Eq.~(\ref{eq:Hamiltonian}) can be made by noticing first that for both models (DM and HM) the defect cavities are roughly equivalent and then have the same resonance frequency, i.e., $f_{nm} = f_{mn} = f_{0}$. Second, the hopping frequency depends on the distance between defects since the solutions in the PnC are evanescent, then for the diagonal and horizontal interferometers those hoppings are expected to be different [see Figs.~\ref{fig:TBmodels} c) and d)]. In addition, the hopping frequencies between nearest neighbors in the bulk defects can be considered the same, i.e., $\nu_{i,j} = \nu$, while the ones at the left and right of the interferometers can be considered as $\nu'$. Thus, the TB model of Eq.~(\ref{eq:Hamiltonian}) only requires the knowledge of the parameters $f_{0}$, $\nu$, and $\nu'$ for each interferometer. Now since $f_{0}$ determines the center of the band, from the experimental spectrum we obtain that $f_{0} \approx 432.0\,  (430.7)$ kHz for the H (D) interferometer (see Fig.~\ref{fig:CRAWsExperiment} a)). Meanwhile, the nearest-neighbor frequency hopping determines the width of the band associated with the defects: the bandwidth is equal to four times the hopping frequency~\cite{MarkosBook}. 

On the one hand, from the experimental frequency spectrum for the horizontal model, the bandwidth is approximately 48.08 kHz and then $\nu \approx 12.0$ kHz with $\nu = \nu'$ as the distance between the defects is the same. For the diagonal model, on the other hand, $\nu \neq \nu'$ since the distance between the defects for this model is different. Furthermore, from the bandwidth of the miniband observed in the frequency spectrum ($\sim 16.0$ kHz), $\nu' \approx 4.2$ Hz and $\nu$ can be estimated as 57.0 kHz. The latter could be obtained from the full experimental spectrum of the diagonal model that unfortunately we do not have, as it is necessary to experimentally measure the power at each cavity.

%
\begin{figure}
\centering
\includegraphics[width=0.8\linewidth]{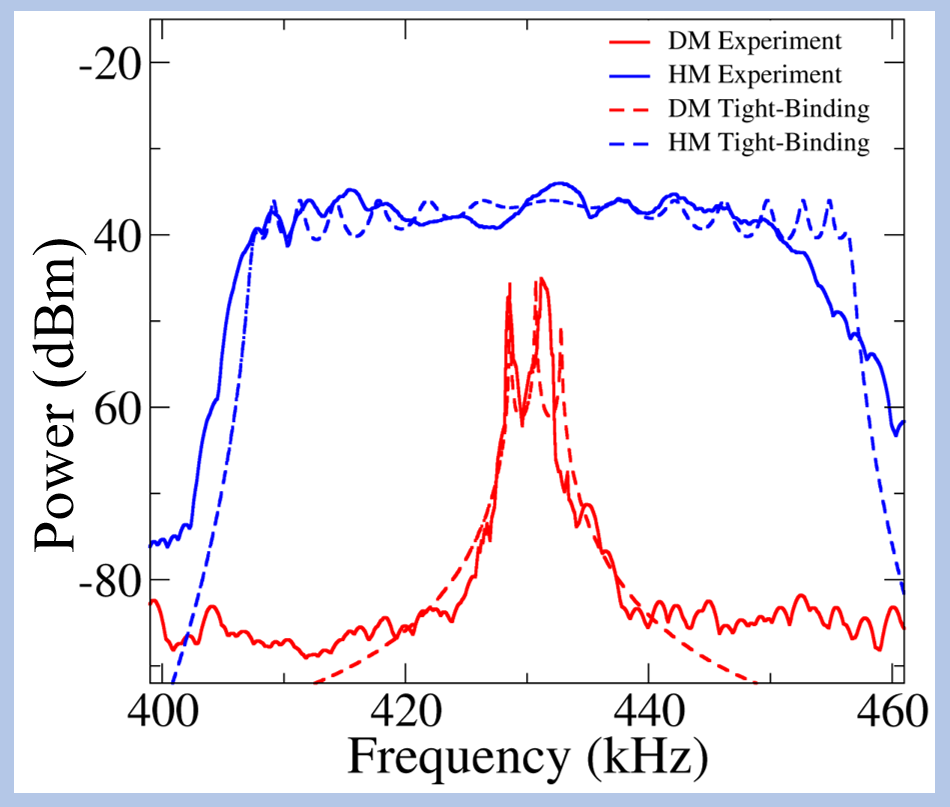}
\caption{(Color online) Power as a function of frequency for the horizontal and diagonal models. Red and blue continuous lines are the experimental results for the DM and HM, respectively. The red and blue dashed lines are the results for the respective tight-binding model of equation~(\ref{eq:Hamiltonian}) together with equations~(\ref{eq:Smatrix}) - (\ref{eq:amplitudes}).}
\label{fig:PowerCRAWsComparison}
\end{figure}

In order to obtain the $S$-matrix for each H (D) model, the coupling strength to the exterior, $\gamma^{\mathrm{L(R)}}$, at the left (L) and right (R) sides has to be determined. For simplicity, we consider that for each model this coupling strength is the same, that is $\gamma^{\mathrm{L(R)}} = \gamma$. \textbf{Here, there are two main leaking mechanisms of each interferometer to the exterior. One is coming from the input and output defects and the other one coming from the interferometer arms. Notice that the leaking from the diagonal interferometer is much smaller than that from the horizontal interferometer since the defects forming the arms of the H-based interferometer are closer to the borders of the PnC. Then, the coupling strength $\gamma$ is fixed by adjusting the width of the resonances within the band associated with each interferometer and corresponds to $\gamma = 10.0\, (0.005)$ kHz for the horizontal (diagonal) interferometer. Table~\ref{tab:parameters} summarizes the TB and $S$-matrix parameters for each interferometer.}

Figure~\ref{fig:CRAWsTB} displays the normalized transmission as a function of frequency for the diagonal and horizontal models in red and blue lines, respectively, obtained through the TB model of Eq.~(\ref{eq:Hamiltonian}) together with the $S$ matrix of (\ref{eq:Smatrix}). The parameters for each model are given in Table~\ref{tab:parameters}. \textbf{The bandwidth is equal to four times the hopping frequency, then, from the experimental spectrum for the HM, this bandwidth is approximately 48.08 kHz. For the DM we only have an experimental spectrum from 400 to 460 kHz which covers only the of interest miniband of approximately 16.8 kHz shown in Fig. 2 a). This allows us to obtain the hopping frequency of 4.2 kHz. To obtain the hopping frequency nu experimentally, the full bandwidth is needed. However, by using TB calculations the value of $\nu = 57.0$ kHz was obtained, which could be slightly different for an experimental spectra of 300 - 500 kHz, for example.} The inset in Figure~\ref{fig:CRAWsTB} provides the normalized transmission on a logarithmic scale. For the horizontal model (blue line), the resonant frequencies around 430 kHz align with the in-gap resonances obtained from the experiment for this model. Conversely, for the diagonal model (red line), it is noteworthy that due to the weak coupling to the outside, the resonances observed are narrower than those obtained for the horizontal model, which is in complete agreement with the experimental spectra and bandstructure results described above.

Finally, Figure~\ref{fig:PowerCRAWsComparison} summarizes the behavior of the transmitted intensity (power) as a function of frequency for both DM and HM interferometers experimentally obtained using ultrasonic spectroscopy (continuous lines), and through the tight-binding model of equation~(\ref{eq:Hamiltonian}) together with equations~(\ref{eq:Smatrix}) - (\ref{eq:amplitudes}) (dashed lines). The plots have been overlapped in the same frequency range for a better comparison, and the intensities were adjusted to demonstrate the agreement and tendencies of the acoustic response of the designed interferometers for each experimental and theoretical analysis. \textbf{It is convenient to appoint out the obtained TB parameters and the coupling strength $\gamma$ are used in Eqs. (3) - (5) and then into the S-matrix of Eq. (2) which now is only a function of frequency $f$. The transmitted power as a function of $f$ is obtained from the S-matrix element $|t|$ as $10Log_{10} (|t|)$ varying the frequency range from 400 to 460 kHz, as shown in Fig. 7 for each interferometer. This gives the TB acoustic power similar to the experimental ones, as previously described}


In the experimental results, slight fluctuations due to errors in locating the cylinders on the respective arrays can be observed. Additionally, a certain degree of power loss is expected, as seen in Figure~\ref{fig:CRAWsExperiment}. These effects can be accounted for in the tight-binding model by introducing a certain degree of disorder, such as in the frequency resonance of the defect cavities, and incorporating an imaginary part in the frequency resonance of the impurity to consider the effect of absorption. 
However, even without the inclusion of these effects, the tight-binding model presented here notably captures the essential features of the horizontal and diagonal models, as observed in both figures~\ref{fig:CRAWsTB} and \ref{fig:PowerCRAWsComparison}, showing high concordance with the other methods of analysis discussed here. When comparing the results with FEM calculations using COMSOL, a left-shifting of 3.5 kHz can be observed. This shifting is attributed to the finite size of the simulated crystal in the FEM, which differs from the idealized infinite crystal assumption in the tight-binding model, and has been previously observed by our group~\cite{Reyes2020}. Figure~\ref {fig:PowerCRAWsComparison}, which shows consistent results, leads to propose the possibility of predicting resonant peaks in CRAWs forming defects-based Mach-Zehnder-type acoustic spectrometers with different and optimized configurations for the selective transmission of ultrasonic waves.

\section{Conclusion}
In conclusion, this study successfully adapted the tight-binding model (TBM) to effectively describe the acoustic response of defect-based acoustic interferometer-like designs, specifically Coupled Resonator Acoustic Waveguides (CRAWs) in two-dimensional phononic crystals (PnCs). The phononic crystal was fabricated with steel cylinders arranged in a square lattice immersed within a water matrix, and defects were induced by removing cylinders to create Mach-Zehnder-like (MZ) defects-based interferometers. Two MZ-type defect-based acoustic interferometers were constructed, with arms oriented horizontally or diagonally, and their transmission features were thoroughly characterized through experimental ultrasonic spectroscopy. By employing the tight-binding approximation, each defect was treated as a resonator coupled to its neighboring ones, which allowed for accurate predictions of the acoustic response within the CRAWs system. The obtained results were compared with experimental ultrasonic spectroscopy data, FEM simulations, and bandstructure calculations. Remarkably, our results exhibited outstanding agreement with the experimental and simulation data, providing robust evidence of the reliability and validity of the proposed TBM description, an approach of solid-state electronics principles successfully proposed for underwater acoustic devices performance. The achieved comprehensive match between the theoretical predictions and the experimental observations holds crucial significance, as it offers an essential tool for precisely predicting and optimizing resonant modes supported by defect arrays within the phononic crystal. Finally, the TBM formulation applied here for describing the CRAWs can straightforwardly be extended to study wave propagation in three-dimensional configurations~\cite{Krushynska2021} which will be a subject of research for future works.


\section{Experimental Section}
To observe the interferometric behavior of the guided waves occurring in the designed defects-based a\-cou\-stic interferometers, the experimental setup illustrated in Figure~\ref{fig:ExpModels} c) is used. The excitation of waves is provided by a Frequency Generator equipment (Teledyne Lecroy, Wave Station-2012) using an unfocused immersion transducer (Olympus V301, 0.5 MHz) as the emitter element. Another transducer, connected to a Spectrum Analyzer (Tektronix-MDO 3024b), serves as the receiver which records the acoustic response. For the analysis, both transducers are positioned face to face at a distance of 50 mm, with the PnC placed in between. The three elements are then immersed in a tank filled with 40 liters of deionized water for measurements at room temperature (22 ${}^{\circ}_{}\mathrm{C}$). The utilized PnC has been thoroughly characterized and possesses a bandgap approximately spanning from 380 to 480 kHz~\cite{Walker2020}. The acoustic response for both models DM and HM, along with the perfect crystal, is recorded in the frequency range from 390 to 475 kHz, corresponding to the bandgap of the perfect crystal.

The experimental pressure maps were obtained using the same set as in Figure~\ref{fig:ExpModels} c), however, the transducer used as the receiver was changed by a needle hydrophone (Muller) of 0.5 mm diameter connected to an amplifier. The hydrophone was moved using a translation stage to record the pressure map covering an area of 40x60 mm in front of the HM interferometer (see Figure ~\ref{fig:CRAWsExperiment} c)), recording the transmitted signal each 2 mm in both directions, starting 2 mm away from the phononic structure. A group of 600 data points were used to generate the experimental pressure map for the frequency of interest.

\medskip
\textbf{Acknowledgements} \par 
DME acknowledges CONAHCyT for the grant to develop Ph.D. studies at UAEMex. A.M.M-A. acknowledges financial support from CONAHCyT under the program ``Estancias Posdoctorales por M\'exico 2022". DR Thanks to UAEMex for the partial financial support through project 6753/2022CIB. AN acknowledges the support from the Distinguished Professorship start-up funds from UESTC and the Ministry of Science and Technology of China (MOST) International Collaboration Grant No. 2022YFE0129000 entitled “Cavity Acoustotodynamics for nonreciprocal wave propagation". RAMS was supported by DGAPA-UNAM under project PAPIIT-IN111021.

\medskip

%

\end{document}